\documentclass[twocolumn,prb]{revtex4}
\usepackage{multirow}
\usepackage{graphicx}
\usepackage{amsmath}
\usepackage{enumerate}
\usepackage{epstopdf}
\usepackage{times}
\usepackage{color}
\usepackage{bm}

\usepackage[colorlinks,bookmarks=false,citecolor=blue,linkcolor=red,urlcolor=blue]{hyperref}

\begin{document}
\title{Spin-$1/2$ Heisenberg antiferromanget on kagome: a $Z_2$ spin liquid with fermionic spinons}
\author{Zhihao Hao}
\affiliation{Department of Physics and Astronomy, University of Waterloo, Waterloo, ON, N2L 3G1, Canada}
\author{Oleg Tchernyshyov}
\affiliation{Department of Physics and Astronomy, Johns Hopkins University, Baltimore, Maryland 21218, U.S.A.}
\begin{abstract}
Motivated by recent numerical and experimental studies of the spin-$1/2$ Heisenberg antiferromagnet on kagome, we formulate a many-body model for fermionic spinons introduced by us earlier [Phys. Rev. Lett. 103, 187203 (2009)]. The spinons interact with an emergent $U(1)$ gauge field and experience strong short-range attraction in the $S=0$ channel. The ground state of the model is generically a $Z_2$ liquid. We calculate the edge of the two-spinon continuum and compare the theory to the slave-fermion approach to the Heisenberg model.
\end{abstract}
\maketitle

\section{Introduction}
\begin{figure}
\centering
\includegraphics[width=0.95\columnwidth]{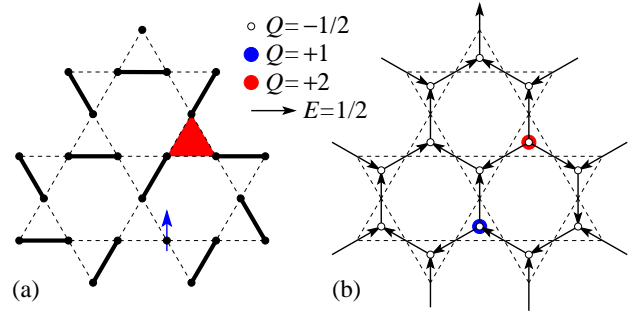}
\caption{(a) The kagome lattice with dimers (black thick bonds), a defect triangle (the red triangle) and a spinon (the blue spin). The corresponding arrow representation is displayed in part (b). }\label{fig:kagome}
\end{figure}

The spin-$1/2$ Heisenberg antiferromagnet on kagome (Fig.~\ref{fig:kagome}) has been extensively studied for over two decades. \cite{PhysRevLett.62.2405, LMM} Combining geometrical frustration and strong quantum fluctuation, the model is expected to host unconventional magnetic orders as well as fractionalized excitations. Several Cu$^{2+}$ based kagome materials have been synthesized. Among them,  one of the most promising realization \cite{helton:107204} of the model is the herbertsmithite, ZnCu$_3$OH$_6$Cl$_2$. Bulk susceptibility \cite{helton:107204} as well as nuclear magnetic resonance (NMR) studies \cite{imai:077203,olariu:087202} of the herbertsmithite conclude that the ground state is a gapless spin liquid. Inelastic neutron scattering study \cite{Han2012} has revealed broad diffusive structure factor instead of sharp features associated with magnons. Relevant perturbations are anisotropic interactions, especially the Dzyaloshinski-Moryia (DM) term,\cite{zorko:026405} and the presence of paramagnetic impurities.\cite{doi:10.1021/ja1070398}

On the theory side, pioneering two-dimensional density matrix renormalization group (DMRG) calculations by several groups \cite{jiang:117203,Yan03062011,PhysRevLett.109.067201,Jiang2012} have established that the ground state of the model is a $Z_2$ spin liquid with a finite spin gap of approximately $0.1J$, where $J$ is the strength of the exchange coupling. Although the finite spin gap seems to contradict experiments, it is hoped that the inconsistency can be resolved by taken into account the additional perturbations in materials. While the effect of DM interaction can certainly be tackled by powerful numerical techniques such as two dimensional DMRG, the effect of Cu substitutions can be tricky to obtain due to the size constraint of the finite cluster.

An analytical framework is thus needed to bridge the gap. Ideally, such a theory should yield a $Z_2$ liquid ground state and incorporate the effects of lattice defects and other perturbations. One possible route is offered by slave-particle approaches, in which spin variables $\mathbf S = (S^x, S^y, S^z)$ are expressed in terms of new fictitious particles:
\begin{equation}
\mathbf S = \frac{1}{2} a_\alpha^\dagger \bm \sigma_{\alpha\beta} a_\beta,
\label{eq:slave-spin}
\end{equation}
where the flavor index $\alpha = \pm 1/2$ labels particles with up and down spins, and $\bm \sigma_{\alpha\beta}$ is the triplet of Pauli matrices. For either bosonic or fermionic ladder operators $a^\dagger_\alpha$ and $a_\alpha$, the variables (\ref{eq:slave-spin}) satisfy the spin commutation relations. The spin length $S$ is set to 1/2 by fixing the net number of particles on a site, $a^\dagger_\alpha a_\alpha = 1$. The binary exchange interaction of the Heisenberg model translates into quartic interactions between the new particles, so that the new problem is not easier to solve than the original one. Solutions are usually based on a mean-field approximation, in which the constraint on the particle numbers is satisfied only on average, $\langle a^\dagger_\alpha a_\alpha \rangle = 1$. The mean-field solution is justified if the number of particle flavors $N$ is increased from two to infinity. For a finite $N$, and particularly for the physical case $N=2$, enforcing the particle-number constraint requires going beyond a mean-field treatment.

For the Heisenberg antiferromagnet on kagome, the bosonic route yields a $Z_2$-liquid ground state with magnetic excitations carrying spin 1/2, known as spinons. Like the particles from which spins are constructed, the spin excitations exhibit the Bose statistics.\cite{PhysRevB.45.12377} Adding DM interactions of sufficient strength induces a quantum phase transition from a gapped $Z_2$ spin liquid state to a gapless state with long-range magnetic order.\cite{PhysRevB.81.144432,PhysRevB.81.064428}  The slave particles can also be fermions \cite{PhysRevB.63.014413,PhysRevLett.98.117205,PhysRevB.77.224413,PhysRevB.83.224413}. Hastings \cite{PhysRevB.63.014413} classified possible states with broken symmetries, including valence-bond solids and liquids starting from a parent state with relativistic spinons as low energy excitations. Later works \cite{PhysRevLett.98.117205,PhysRevB.77.224413,PhysRevB.84.020407} concluded that the parent state is the ground state by optimizing numerically the Gutzwiller-projected mean-field solutions using variational Monte Carlo. Based on an projective symmetry group analysis,  Lu \emph{et al}\cite{PhysRevB.83.224413} suggest that one of the candidate ground states is similar to the $Z_2$ liquid state found in DMRG calculations.

It is clear that the choice of particle statistics at the starting point (\ref{eq:slave-spin}) determines the statistics of spin excitations, at least at the level of the mean-field approximation. It is therefore desirable to determine the statistics of excitations at the outset by some other means. In our previous work,\cite{PhysRevLett.103.187203} we presented arguments that the Heisenberg antiferromagnet on kagome has fermionic spinons. To do so, we constructed states with two spin-1/2 solitons that resemble closely spinons of the one-dimensional antiferromagnet on a sawtooth chain \cite{PhysRevB.53.6401,PhysRevB.53.6393}. We showed that the wavefunction is antisymmetric under the exchange of two spinons, hence fermionic statistics. This calculation was performed on a tree version of the kagome lattice.\cite{PhysRevB.48.13647} The deformation of the lattice was necessary to control the number of spinons. Strong attraction, mediated by exchange interaction, binds two spinons into a small pair with spin zero, which manifests itself as a defect---a triangle lacking a valence bond. A kagome lattice proper has a finite concentration of such defects: one in four triangles. This counting yields one spinon per unit cell on kagome. The main goal of this paper is to build a many-body model of interacting spinons of this kind.

Our approach is variational in essence. One part of the variational basis consists of ``dimer-covering'' states where all spins form singlets, ``dimers'', with one of their nearest neighbors. The wave function of such a dimer between site $i$ and site $j$ is:
\begin{equation}\label{eq:dimer}
|ij\rangle_b = \frac{1}{\sqrt{2}}(|i\uparrow, j\downarrow\rangle-|j\uparrow, i\downarrow\rangle).
\end{equation}
To prevent a sign ambiguity, the dimer state $|ij\rangle_b$ is represented with an arrow pointing from site $i$ to site $j$. This is the bosonic convention \cite{PhysRevB.40.7133,PhysRevB.81.214413} for dimer wave function as we can write it by using the Schwinger bosons to represent spins:
\begin{equation}
|ij\rangle_b =  2^{-1/2}
	(b_{i\uparrow}^\dagger b_{j\downarrow}^\dagger - b_{i\downarrow}^\dagger b_{j\uparrow}^\dagger)|0\rangle
	= -|ji\rangle_b
\end{equation}
The reason to includes such states is as follows. The Heisenberg Hamiltonian on a kagome lattice can be rewritten in terms of the total spins of individual triangles $\mathbf S_\Delta$:
\begin{equation}\label{eq:hami}
H=J\sum_{\langle ij\rangle}\mathbf{S}_i\cdot\mathbf{S}_j=\frac{J}{2}\sum_{\Delta}\left(\mathbf S_\Delta^2-\frac{9}{4}\right).
\end{equation}
For spins of length $S=1/2$ the energy would be minimized if every triangle had the lowest possible spin, $S_\Delta = 1/2$. This can be achieved if two of the three spins on every triangle form a singlet (a quantum dimer), leaving the third spin to form a singlet on another triangle, Fig.~\ref{fig:kagome}. This program can be realized for a one-dimensional analog of kagome, the sawtooth chain,\cite{PhysRevB.53.6393, PhysRevB.53.6401} which has two dimerized ground states. Unfortunately, the trick does not work on kagome, where one quarter of triangles lack a dimer.\cite{PhysRevLett.62.2405} In our previous work,\cite{PhysRevLett.103.187203} we demonstrated that these defect triangles are bound states of two fermionic spinons with total spin 0. This translates into two spinons for every four triangles, or one spinon per unit cell of kagome. Although this is different from the standard slave-fermion approach, in which there are three spinons per unit cell (one per site), the two pictures are closely related as we discuss in later sections. Our spinons live on the honeycomb lattice formed by the centers of kagome triangles. Their motion is strongly constrained by the presence of quantum dimers. These constraints can be described in terms of an emergent compact $U(1)$ gauge field. In addition to interacting with the gauge field in the usual manner, our spinons experience strong short-range attraction to each other in the $S=0$ channel. This is captured by an on-site negative Hubbard $U$ interaction between spinons. Because attraction between fermions generally induces Cooper pairing, the low energy theory of our model is a compact $U(1)$ gauge theory interacting with a charge-2 Higgs field, which has a $Z_2$ liquid ground state. \cite{PhysRevD.19.3682,Bhanot1981357} Excitations are fluxes of the $Z_2$ gauge field (visons) and deconfined fermionic quasiparticles (spinons).

The rest of the paper is organized as follows. In Sec \ref{husimi}, we briefly review the solution of two spinons on the Husimi Cactus, a tree of corner sharing triangles. We then review the arrow representation of dimer coverings and extend it to include spinons. The model is written down in Sec \ref{model} . We spent the next few sections to solve the model in various limits. We first study the model in $U=0$ limit in Sec \ref{zerolimit} and obtain two saddle point solutions: the zero and $\pi$ flux phases. For $U$ large enough, both saddle points host a $Z_2$ liquid phase as the ground state. Since it is more natural to realize small spin gap in the $\pi$ flux phase, we identify it as the saddle point that better describes the kagome. We calculate the lower edge of two-spinon continuum of the $Z_2$ liquid originated from the $\pi$ flux phase. Based on previous works, we discuss the nature of the finite temperature phase transition to the $Z_2$ liquid phase.  We conclude the paper with discussion of the relation between our works and previous works, possible experimental and numerical comparisons as well as future directions.

\section{The construction of the model}\label{mc}
In this section, we build the model. We first motivate the elements of the model by reviewing previous works. We then write down the model and explain its basic properties.
\subsection{An isolated defect triangle on the Husimi cactus}\label{husimi}
Before we describe the many body model, it is instructive to briefly review the problem of an isolated defect triangle on the Husimi cactus \cite{PhysRevLett.103.187203,PhysRevB.81.214445}.
\begin{figure}
\centering
\includegraphics[width=0.95\columnwidth]{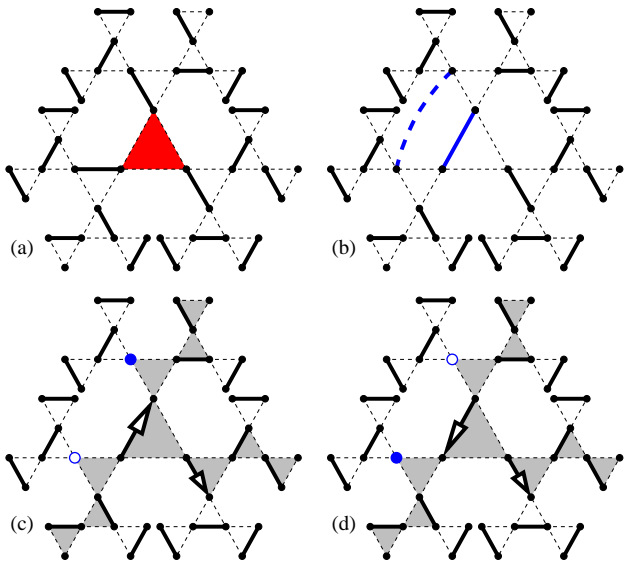}
\caption{(a) An isolated defect triangle at the center of the Husimi Cactus. (b) The defect triangle is broken up into two spinons connected by a long range singlet (the blue dashed line). (c),(d) The exchange of two spinons reverse the ``direction'' of one singlet.}\label{fig:cactus}
\end{figure}

The Husimi cactus (Fig \ref{fig:cactus}) is the Cayley tree of corner sharing triangles. It reproduces the local geometry of kagome without the presence of loops. The ground state of the Hamiltonian \eqref{eq:hami} is any dimer covering state since all triangles are vacuum triangles of \emph{lowest} possible energy. An defect triangle can be isolated at the center of the Cactus. We label the state $|0,0,0\rangle$.

By applying the exchange Hamiltonian, we map out the full Hilbert space, $\Gamma_2$ ,that $|0,0,0\rangle$ belongs to. Generally, a state in $\Gamma_2$ is characterized by the position of two mobile spin $1/2$ particles, spinons, connected by a long range singlet. The two spinons move on three one dimensional trails $x$, $y$ and $z$ \cite{PhysRevB.81.214445} connected to the center of the cactus. Such a state is labeled as $|x,y,z\rangle$ with the constraint that $xyz=0$. The dimers constrain the motion of spinons by determining the three one dimensional paths.

We project the exchange Hamiltonian \eqref{eq:hami} to the $\Gamma_2$. For example, we have:
\begin{equation}\label{eq:effhami}
\begin{aligned}
H|x,y,0\rangle=&-t\left(|x+1,y,0\rangle+|x-1,y,0\rangle+|x,y+1,0\rangle\right.\\ &\left.+|x,y-1,0\rangle\right)-U\delta_{x,0}\delta_{y,0}|0,0,0\rangle
\end{aligned}
\end{equation}
where $t=J/2$ and $U=3J/4$. The spinons tend to delocalize to gain kinetic energy. On the other hand, if they form a nearest neighbor singlet adjacent to a defect triangle, the potential energy of the system is lowered by $U$. We obtained the spectrum of the two spinons with total $S=0$. \cite{PhysRevLett.103.187203,PhysRevB.81.214445} In addition to the two particle continuum, the two spinons can form a bound state gaining energy $0.06J$.

The cactus also provides us with an opportunity to determine the statistics of spinons by performing their adiabatic exchange.\cite{PhysRevLett.103.187203, PhysRevB.81.214445} After an exchange, the wave function acquires a negative sign because an odd number of dimers change from $|ij\rangle_b$ to $|ji\rangle_b = -|ij\rangle_b$. The fermionic statistics of spinons comes about in a rather tortuous way in the bosonic convention for spin singlets. It follows much more naturally if we use a fermionic convention\cite{PhysRevB.40.7133,PhysRevB.81.214413} instead:
\begin{equation}\label{eq:fermiconvention}
|ij\rangle_f =  2^{-1/2}
	(a_{i\uparrow}^\dagger a_{j\downarrow}^\dagger - a_{i\downarrow}^\dagger a_{j\uparrow}^\dagger)|0\rangle
	= |ji\rangle_f,
\end{equation}
where $a_{i\sigma}$ creates a fermion with spin $\sigma$ on site $i$. Under this convention, the phase ambiguity is absent as $|ij\rangle_f=|ji\rangle_f$. Spinon motion can be described in either bosonic or fermionic representation of spin singlets (Appendix \ref{saw}).

On kagome lattice, there is a finite concentration of spinons. These fermions have two opposite tendencies: delocalizing to lower kinetic energy and binding into nearest neighbor dimers to gain potential energy. Their motions are also constrained by the dimer configurations. To characterize the constrains, we introduce  the arrow representation.

\subsection{Arrow representation}\label{arrow}
Zeng and Elser \cite{PhysRevB.51.8318}, and later Misguich \emph{et al.},\cite{PhysRevLett.89.137202, misguich:184424} used an arrow representation for dimer coverings on kagome. The arrows live on links of a dual honeycomb lattice, whose sites are centers of kagome triangles. When a quantum dimer is present on a triangle, two arrows point into this triangle through the ends of the dimer. Thus a triangle containing a dimer has two arrows pointing in and one out. A defect triangle has three arrows all pointing out. A spinon (of the antikink type\cite{PhysRevLett.103.187203, PhysRevB.81.214445}) has one arrow in and two out, Fig.~\ref{fig:kagome}.

In the arrow representation, spinons live on honeycomb sites and move in the direction of arrows only. Thus the spinon in Fig.~\ref{fig:kagome}(b) can only move up or to the left, but not to the right. As a spinon moves across a link, the arrow on that link reverses its direction. These rules preserve the right arrow count (one in, two out) for the moving spinon. In what follows, we use boldface indices $\mathbf{r}$ to represent sites of the honeycomb lattice and italic indices $i$ for kagome sites.

\subsection{Compact $U(1)$ gauge theory}

The arrow representation can be reparametrized as a $U(1)$ gauge theory on the honeycomb lattice. We define a gauge potential $A_{\mathbf r \mathbf r'} = - A_{\mathbf r' \mathbf r}$ on a link connecting neighboring honeycomb sites $\mathbf r$ and $\mathbf r'$. The gauge field is compact in such a way that its wavefunction is antiperiodic: $\psi(A_{\mathbf r \mathbf r'} + 2\pi) = -\psi(A_{\mathbf r \mathbf r'})$. The momentum conjugate to the gauge field, the link electric field $E_{\mathbf r \mathbf r'} \equiv -i \partial/\partial A_{\mathbf r \mathbf r'} = -E_{\mathbf r' \mathbf r}$, takes on half-integer values $\pm 1/2, \pm 3/2, \pm 5/2, \ldots$ We identify arrows on honeycomb links with the smallest (in the absolute sense) values of the electric field, $E_{\mathbf r \mathbf r'} = \pm 1/2$. The electric charge is defined as the electric flux emerging from a site, $Q_{\mathbf r} = \sum_{\mathbf r'} E_{\mathbf r \mathbf r'}$. By this definition, a triangle with a quantum dimer (and thus in a ground state) carries charge $Q_0=-1/2$, which can be regarded as a background charge of the vacuum. A honeycomb site with a spinon has charge $Q = +1/2$. A defect triangle (a spinon pair) has $Q = +3/2$. Subtracting the background charge, we find that spinons carry charge $Q - Q_0 = +1$ and defect triangles (spinon pairs) $Q - Q_0 = +2$.

The constrained dynamics of spinons, including their attraction in the $S=0$ channel, is captured by the following Hamiltonian:
\begin{equation}\label{eq:cpu1}
H = \sum_{\langle \mathbf r \mathbf r' \rangle}
	\left(
		\frac{E_{\mathbf r \mathbf r'}^2}{2\epsilon}
		- t \sum_\sigma a^\dagger_{\mathbf r\sigma} e^{i A_{\mathbf r \mathbf r'}} a_{\mathbf r'\sigma}
	\right)
	- U \sum_{\mathbf r} n_{\mathbf r \uparrow} n_{\mathbf r \downarrow}.
\end{equation}
The first term in the Hamiltonian is a compact $U(1)$ gauge theory with fermions of charge 1. The $E_{\mathbf r \mathbf r'}^2/2\epsilon$ term removes states with electric fields exceeding the minimal absolute value of 1/2. The gauge factor $e^{i A_{\mathbf r \mathbf r'}}$ increments the electric field $E_{\mathbf r \mathbf r'}$ by 1 as a fermion moves from $\mathbf r'$ to $\mathbf r$. Note the absence of the standard magnetic flux term $\cos{\Phi}$, where the flux $\Phi$ through a hexagonal plaquette is the lattice analog of a line integral of the gauge field $\oint \mathbf A \cdot d\mathbf r$.

Even in the absence of fermion self-interaction (the Hubbard $U$ term) the theory \eqref{eq:cpu1} is not exactly solvable. In the unphysical limit of large $\epsilon$, the Hamiltonian becomes diagonal in the gauge fields $A_{\mathbf r \mathbf r'}$, which have no dynamics of their own and merely set a magnetic flux for the fermions. In this limit the standard slave-fermion mean-field approximation becomes exact and the ground state is found by finding a magnetic flux background yielding the lowest kinetic energy for fermions. A state with well-defined fluxes, and thus gauge variables $A_{\mathbf r \mathbf r'}$, is clearly unphysical: it has infinitely large fluctuations of the electric fields $E_{\mathbf r \mathbf r'}$. In the physical small-$\epsilon$ limit, $E_{\mathbf r \mathbf r'}$ is limited to $\pm 1/2$ and gauge variables $A_{\mathbf r \mathbf r'}$ undergo strong fluctuations. We need to write the theory in a more economical form in order to make further progress.

\subsection{Pseudospin representation}\label{model}

An alternative approach is to represent the arrows on links as pseudospins of length $s=1/2$. Using a bipartite nature of the honeycomb lattice, we define $s^z_{\mathbf r \mathbf r'} = s^z_{\mathbf r' \mathbf r} = +1/2$ if the arrow points from sublattice A to sublattice B. The number of spinons on site $\mathbf r$ is related to the pseudospins on the adjoining links:
\begin{equation}\label{eq:gauss}
n_{\mathbf r}
	= s + (-1)^{\mathbf r} \sum_{\mathbf r'}s_{\mathbf r \mathbf r'}^{z},
\end{equation}
with the symbolic notation $(-1)^{\mathbf r} = +1$ if $\mathbf r$ is on sublattice A and $-1$ if $\mathbf r$ is on sublattice B.

%\subsection{The model}\label{model}
In the pseudospinon representation, the Hamiltonian is
\begin{equation}\label{eq:model}
H = - t\sum_{\mathbf r \in A} \sum_{\mathbf r' \in B} \sum_ \sigma
	(a^\dagger_{\mathbf r \sigma} s^+_{\mathbf r \mathbf r'} a_{\mathbf r' \sigma} + \mathrm{H.c.})
	- U \sum_{\mathbf{r}} n_{\mathbf r \uparrow} n_{\mathbf r \downarrow}.
% H=-t\sum_{\langle\mathbf{r}_1\mathbf{r}_2\rangle}\left(a_{\mathbf{r}_1\sigma}^\dagger(A)S^{+}_{\mathbf{r}_1\mathbf{r}_2}a_{\mathbf{r}_2\sigma}(B)+h.c\right)-U\sum_{\mathbf{r}}n_{\mathbf{r}\uparrow}n_{\mathbf{r}\downarrow}.
\end{equation}
The coupling constants $t$ and $U$ are both of the order of $J$, the only energy scale in the Heisenberg model.  A crude estimate based on our study of spinons on a Husimi cactus yields $t = J/2$ and $U = 3J/4$. The number of spinons is 1 per unit cell.

Local constraints \eqref{eq:gauss} give rise to a compact $U(1)$ gauge symmetry. To make it manifest, we write down the Lagrangian for pseudospin variables expressed in terms of angular variables, $s^z = s\cos{\theta}$, $s^\pm = s\sin\theta e^{\pm i \phi}$. A path integral for this theory can be written as
\begin{equation}\label{eq:pathintegral}
\int DA^0 \, D\phi \, D\theta \, Da^\dagger \, Da \,
	\exp{\left( i \int L \, dt \right)},
%\int D(\phi_{\mathbf{r}_1\mathbf{r}_2}(t)\theta_{\mathbf{r}_1\mathbf{r}_2}(t)a_{\mathbf{r}\sigma}^\dagger(t)a_{\mathbf{r}\sigma}(t))\exp\left(i\int dt L\right)
\end{equation}
where
\begin{subequations}
\label{eq:lagrangian}
\begin{eqnarray}
L &=& s \sum_{\langle \mathbf r \mathbf r' \rangle}
		(\cos{\theta_{\mathbf r \mathbf r'}} - 1)
		\dot{\phi}_{\mathbf r \mathbf r'}
	+ i \sum_{\mathbf r, \sigma} a^\dagger_{\mathbf r \sigma} \dot{a}_{\mathbf r \sigma}
\label{eq:lagrangian-a}
\\
	&&+ \sum_{\mathbf r} A^0_{\mathbf r} \left[
			n_{\mathbf r} - s - (-1)^{\mathbf r} \sum_{\mathbf r'} s^z_{\mathbf r \mathbf r'}
		\right]
\label{eq:lagrangian-b}
\\
		&&- H.
\label{eq:lagrangian-c}
\end{eqnarray}
\end{subequations}
The first line (\ref{eq:lagrangian-a}) contains the standard kinetic terms for spins and nonrelativistic fermions; the next term (\ref{eq:lagrangian-b}) enforces the local constraints. The Lagrangian is invariant under $U(1)$ gauge transformations,
\begin{subequations}\label{eq:gauge}
\begin{eqnarray}
a_{\mathbf{r}\sigma} &\to &
	a_{\mathbf{r}\sigma}\exp(i\lambda_{\mathbf{r}}),\\
\phi_{\mathbf r \mathbf r'} &\to &
	\phi_{\mathbf r \mathbf r'} + (-1)^{\mathbf r}(\lambda_{\mathbf r} - \lambda_{\mathbf r'}),\\
A_{\mathbf{r}}^0 &\to &
	A_{\mathbf{r}}^0 + \dot{\lambda}_{\mathbf{r}}.
\end{eqnarray}
\end{subequations}
It is evident that quantities $(-1)^{\mathbf r} \phi_{\mathbf r \mathbf r'}$ transform like spatial components of a lattice $U(1)$ gauge field and $A^0_{\mathbf r}$ as its time component. The angular nature of $\phi_{\mathbf r \mathbf r'}$ makes the gauge field compact. \eqref{eq:model} and \eqref{eq:lagrangian} are our main results. The theory will be solved in various limits in the remainder of the paper.

\section{The $U=0$ limit: gauge mean-field theory}\label{zerolimit}
We first solve the theory in the $U=0$ limit. It is still non-trivial since the fermions are interacting with a compact $U(1)$ gauge field. To make progress, we decompose the kinetic energy terms of fermions using the gauge mean-field theory:\cite{PhysRevLett.108.037202}
\begin{equation}\label{eq:gaugemf}
\begin{aligned}
a_{\mathbf r \sigma}^\dagger s^+_{\mathbf r \mathbf r'} a_{\mathbf r' \sigma} &\approx
	\langle s_{\mathbf r \mathbf r'}^+ \rangle
	a_{\mathbf r \sigma}^\dagger a_{\mathbf r' \sigma}
 + s_{\mathbf r \mathbf r'}^+ \langle a_{\mathbf r \sigma}^\dagger a_{\mathbf r' \sigma} \rangle
 \\
	& - \langle s_{\mathbf r \mathbf r'}^+\rangle \langle a_{\mathbf r \sigma}^\dagger a_{\mathbf r' \sigma}\rangle.
\end{aligned}
\end{equation}
The averages $\langle s_{\mathbf r \mathbf r'}^+\rangle = s\sin\theta_{\mathbf r \mathbf r'}\exp(i\phi_{\mathbf r \mathbf r'})$ play the role of hopping matrix elements for spinons, whereas bond averages $\langle a_{\mathbf{r}\sigma}^\dagger a_{\mathbf r' \sigma}\rangle$ create a transverse magnetic field for pseudospins. The spinon kinetic energy is minimized by setting $\theta_{\mathbf r \mathbf r'} = \pi/2$ for every bond. The spinon spectrum is then dependent on azimuthal angles $\phi_{\mathbf r \mathbf r'}$ that determine magnetic fluxes through hexagonal plaquettes. Below we consider flux configurations that do not break the time reversal symmetry (fluxes 0 and $\pi$) and preserve translational symmetry (all fluxes are the same).

\subsection{Zero flux phase}
In the simplest case, magnetic fluxes on all plaquettes are zero. The mean-field Hamiltonian for the fermions is:
\begin{equation}\label{eq:mf0}
H_\mathrm{MF}=-t s \sum_{\langle \mathbf r \mathbf r' \rangle, \sigma}
	(a_{\mathbf r \sigma}^\dagger a_{\mathbf r' \sigma}
	+ \mbox{H. c.}) - \mu \sum_{\mathbf{r}, \sigma}a_{\mathbf{r}\sigma}^\dagger a_{\mathbf{r}\sigma}.
\end{equation}
The chemical potential $\mu$ is adjusted so that the number of fermions is 1 per unit cell. The primitive vectors are $\mathbf{a}_1=\hat{x}$ and $\hat{a}_2=\hat{x}/2+\sqrt{3}\hat{y}/2$.
\begin{figure}
\centering
\includegraphics[width=0.9\columnwidth]{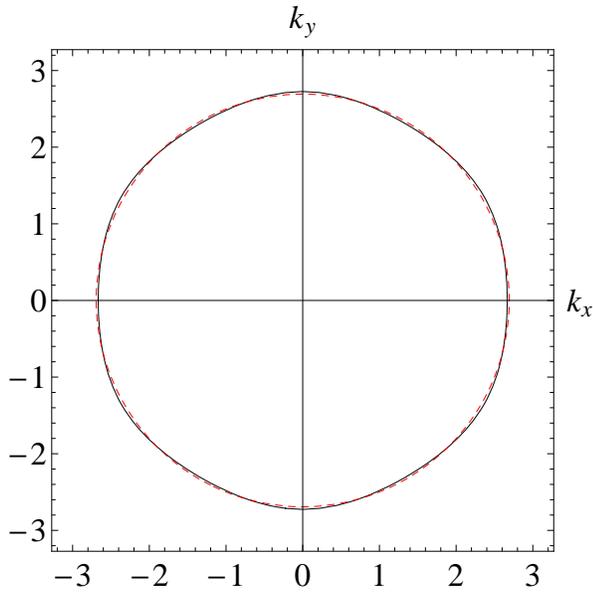}
\caption{The spinon Fermi surface (black) for zero-flux phase. The red-dashed line is obtained by assuming $E = k_f^2/2m$. }\label{fig:zeroF}
\end{figure}
After a Fourier transformation,
\begin{equation}\label{eq:diag0}
H=\sum_{\mathbf{k},\sigma,\tau} \xi_{\tau}(\mathbf{k})c_{\mathbf{k}\sigma\tau}^\dagger c_{\mathbf{k}\sigma\tau}
\end{equation}
where
% $U_{\tau\tau^\prime} (\mathbf{k})c_{\mathbf{k}\sigma\tau^\prime}=a_{\mathbf{k}\sigma\tau}$ and
\begin{equation}\label{eq:spec0}
\xi_{1,2}(\mathbf{k})= \mp tS\sqrt{3+2\cos k_x+4\cos\frac{k_x}{2}\cos\frac{\sqrt{3}k_y}{2}}-\mu.
\end{equation}
% $U(\mathbf{k})$ is an unitary matrix and
The chemical potential $\mu\approx -0.74t$ yields 1 fermion per unit cell. Spinons have a nearly circular Fermi surface (Fig.~\ref{fig:zeroF}) with the Fermi momentum $k_f\approx 2.7$. The effective mass is $m\approx 4.7/t$ and the density of states is $m/2\pi$.

%To check that the solution is consistent, we compute $\langle a_{\mathbf r \sigma}^\dagger a_{\mathbf r' \sigma}\rangle\approx +0.72$. Fermions exert a Zeeman field polarizing the pseudospins along the $x$ direction. The mean-field solution is consistent.

Thanks to the extended Fermi surface, the spinons remain deconfined \cite{PhysRevB.78.085129} even if the compact $U(1)$ gauge fluctuations are taken into account.  We do note that a recent report \cite{2011arXiv1112.6243Z} claims the contrary that compact gauge fluctuations confine spinons with a Fermi surface.
\begin{figure}
\centering
\includegraphics[width=0.95\columnwidth]{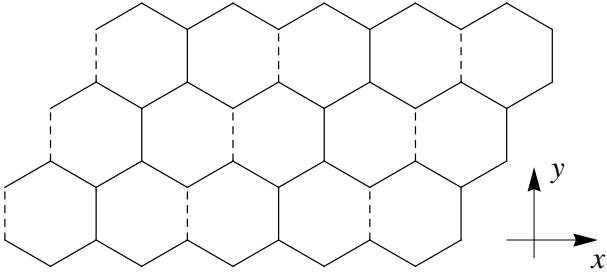}
\caption{$M_{\mathbf r \mathbf r'}=-1$ for the dashed bonds.}\label{fig:pihoney}
\end{figure}

\subsection{The $\pi$ flux phase}
% We consider the saddle point with $\pi$ flux per plaquette.
The unit cell is doubled to include four sites of the honeycomb lattice.  The mean-field Hamiltonian for spinons reads:
\begin{equation}\label{eq:mfp}
H_\mathrm{MF} = - t s\sum_{\langle \mathbf r \mathbf r' \rangle, \sigma}
	M_{\mathbf r \mathbf r'}(a_{\mathbf r \sigma}^\dagger a_{\mathbf r' \sigma} + \mbox{H. c.})
	-\mu \sum_{\mathbf{r}, \sigma}a_{\mathbf{r}\sigma}^\dagger a_{\mathbf{r}\sigma}.
\end{equation}
Here $M_{\mathbf r \mathbf r'}=\pm 1$ for solid (dashed) bonds in Fig.~\ref{fig:pihoney}.
The resulting bands are, in the order of increasing energy,
\begin{subequations}
\label{eq:specp}
\begin{eqnarray}
\xi_{1,2} (\mathbf{k}) = -tS\sqrt{3\pm 2\rho(\mathbf{k})}-\mu,
\\
\xi_{3,4} (\mathbf{k}) = +tS\sqrt{3\mp 2\rho(\mathbf{k})}-\mu,
\end{eqnarray}
\end{subequations}
where $\rho(\mathbf{k})=\sqrt{1+\cos^2k_x-\sin k_x\sin\sqrt{3}k_y}$. The chemical potential $\mu=-\sqrt{3}t/2$ is set at the intersection of bands $\xi_1$ and $\xi_2$, so that only the lowest band $\xi_1$ is filled. The lower two bands touch at two Dirac points with momenta $\mathbf{p}_{1,2}=\pm\left(\pi/2,\pi/2\sqrt{3}\right)$ (Fig \ref{fig:bandpi}). In the vicinity of the two Dirac points $\mathbf{p}_{1,2}$, the Fermi velocity is $t/2\sqrt{2}$.

%The consistency of the mean-field solution can again be checked by computing $\langle a_{\mathbf{r}_1\sigma\alpha}^\dagger a_{\mathbf{r}_2\sigma\beta}\rangle$. Except $\langle a_{\mathbf{r}\sigma0}^\dagger a_{\mathbf{r}+\hat{x}/\sqrt{3},\sigma 1}\rangle$ is negative, all other nearest-neighbor bond averages are positive. These bond averages polarize the pseudospins in the proper directions to produce \eqref{eq:mfp}.
\begin{figure}
\centering
\includegraphics[width=0.9\columnwidth]{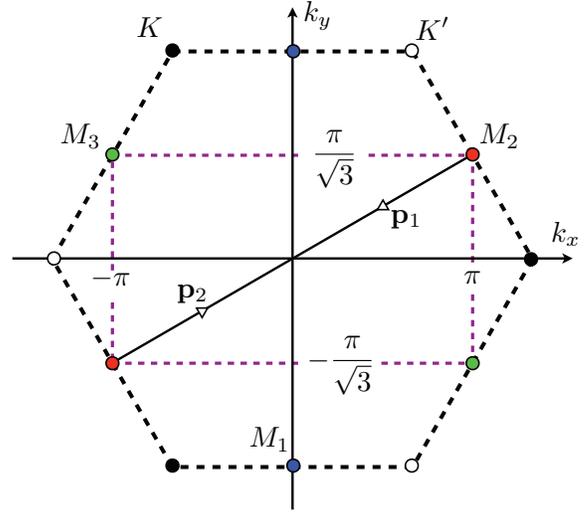}
\caption{The positions of the Fermi points $\mathbf{p}_{1,2}$. The black hexagon is the first Brillouin zone of the honeycomb lattice. The purple rectangle is the Brillouin zone of the $\pi$-flux state. Triangles mark Dirac points $\mathbf p_1$ and $\mathbf p_2$. }\label{fig:points}
\end{figure}

The ``flavor'' number of Dirac fermions is $N_f=4$ considering both the spin and valley degeneracy. In the limit of large $N_f$ \cite{PhysRevB.70.214437}, Dirac fermions are not confined by compact $U(1)$ gauge field. There are conflict reports \cite{Fosco20061843, PhysRevLett.91.171601, PhysRevB.77.045107, PhysRevD.84.014502} on whether $N_f=4$ is above or below the critical number flavors. We will operate under the assumption that spinons are deconfined. Since spinons eventually form Cooper pairs, this assumption is not essential to our conclusion.
%\section{Large $U$ limit}

\section{$U\neq 0$: BCS and the Higgs phase}
We consider now the case of finite $U$. For sufficiently strong attraction, fermions form Cooper pairs and develop a gap around the Fermi surface. We calculate the BCS gap for both 0- and $\pi$-flux states.
\subsection{Zero-flux phase}
The interaction term is decoupled in the usual BCS way:
\begin{equation}\label{eq:bcs0}
-U\sum_{\mathbf{r}}n_{\mathbf{r}\uparrow}n_{\mathbf{r}\downarrow}
	= \frac{2N\Delta^2}{U}
	-\Delta\sum_{\mathbf{k},\tau}
		(c_{\mathbf{k}\tau\uparrow}^\dagger c_{-\mathbf{k}\tau\downarrow}^\dagger + \mathrm{H.c.}),
\end{equation}
where $N$ is the number of unit cells and $\tau = 1,2$ is the band index.
\begin{figure}
  \includegraphics[width=0.48\columnwidth]{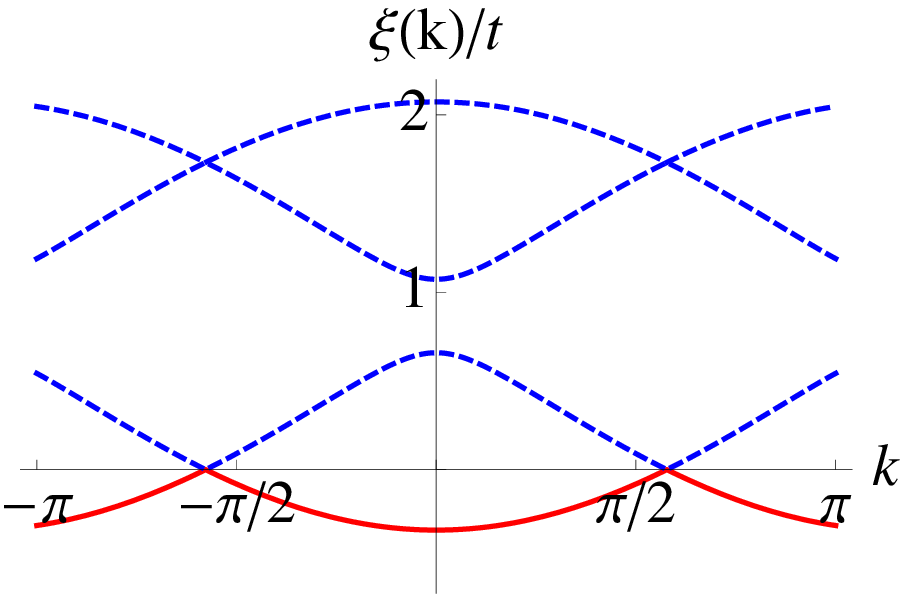}
  \includegraphics[width=0.48\columnwidth]{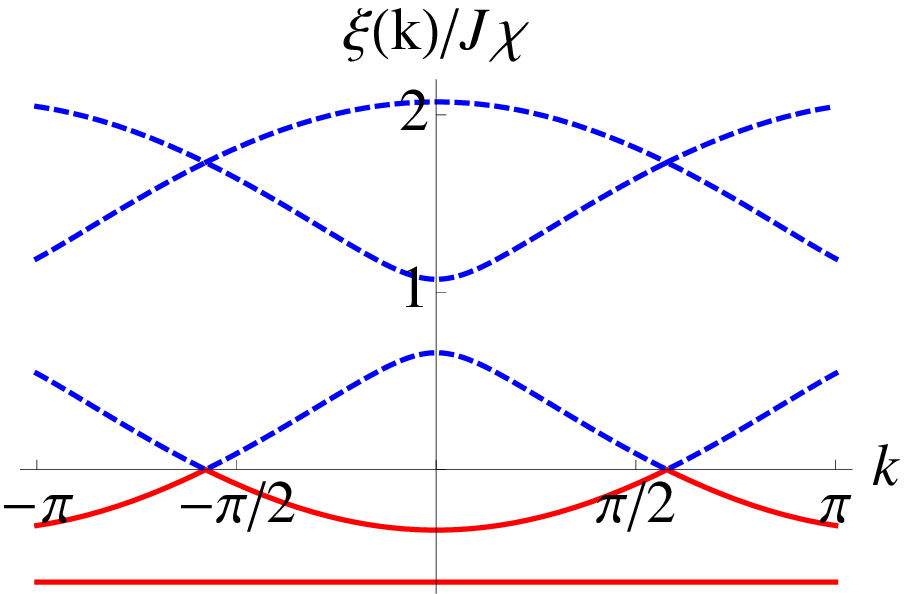}
  \caption{Energy bands of noninteracting spinons with the wavenumber $\mathbf k = (k \cos{\phi}, k \sin{\phi})$ with $\phi = \pi/6$ in the $\pi$-flux phase. Left panel: our model, Eq.~(\ref{eq:specp}). Right panel: slave-fermion theory, Eq.~(\ref{eq:6bands}); the flat band is doubly degenerate. Filled (empty) bands are shown by solid (dashed) lines.}
\label{fig:bandpi}
\end{figure}

A Bogoliubov transformation can be used to diagonalize the Hamiltonian:
\begin{equation}\label{eq:bcsquasi0}
H=\sum_{\mathbf{k},\tau} \varepsilon_\tau(\mathbf k) \, d_{\mathbf{k}\tau}^\dagger d_{\mathbf{k}\tau},
\end{equation}
where $\varepsilon_\tau({\mathbf k}) = \sqrt{\xi_{\tau}^2(\mathbf{k})+\Delta^2}$ is the excitation energy of a Bogoliubov quasiparticle. The self-consistency condition, $\Delta = U \langle a_{\mathbf r \uparrow} a_{\mathbf r \downarrow}\rangle$, yields the gap equation:
\begin{equation}\label{eq:gap0}
1=\frac{1}{4N}\sum_{\mathbf{k},\tau}\frac{U}{\sqrt{\xi_{\tau}^2(\mathbf{k}) + \Delta^2}}.
\end{equation}
The presence of extended Fermi surface implies that $\Delta\neq 0$ for infinitesimal $U$. For the bare couplings, $t = J/2$ and $U=3J/4$, we obtain $\Delta=0.39t$. The spin gap is then $2\Delta = 0.39J$, much larger than the DMRG value $0.1J$. In order to reduce the spin-gap to numerically observed value, large modifications of $U$ or $t$ from their bare values are needed, which is not likely.
\subsection{The $\pi$-flux phase}
Similar calculations can be carried out for the $\pi$-flux phase.
% We use $V=N/2$ to denote the number of unit cells.
The gap equation reads:
\begin{equation}\label{eq:gappi}
1=\frac{1}{4N}\sum_{\mathbf{k},\tau}\frac{U}{\sqrt{\xi_{\tau}(\mathbf{k})^2+\Delta^2}}.
\end{equation}
Due to vanishing density of state around the Dirac points, the gap equation is only satisfied for $U>U_c$ where
\begin{equation}\label{eq:uc}
1=\frac{1}{4N}\sum_{\mathbf{k},\tau}\frac{U_c}{|\xi_{\tau}(\mathbf{k})|}.
\end{equation}
Explicit calculations show $U_c\approx t$. For the bare couplings, $t = J/2$ and $U=3J/4$, the spin gap $2\Delta\approx 0.32J$. While it is still large, relatively small adjustments to parameters can improve the agreement greatly. For example, for an enhanced spinon hopping $t = 0.68J$ and the bare $U=3J/4$, we find the spin gap $2\Delta = 0.1 J$.
% We thus conclude that the $\pi$-flux phase is a better description of spin $1/2$ Heisenberg antiferromagnet on kagome.

\begin{figure}
\centering
\includegraphics[width=0.95\columnwidth]{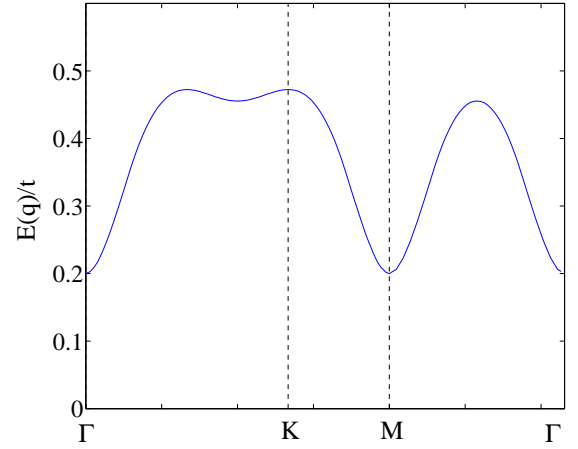}
\caption{The lower edge of the two particle continuum for the $\pi$-flux phase assuming $\Delta=0.1t$.}\label{fig:edge}
\end{figure}

Figure \ref{fig:edge} shows the lower edge of the two-spinon continuum for $\Delta=0.1t$ in the $\pi$-flux phase along high-symmetry directions.  Our theory predicts low energy $S=1$ excitations at both $\Gamma$ and $\mathrm{M}$ points (Fig \ref{fig:points}). This can be understood without explicit calculation. The low energy spinon excitations are of momentum $\mathbf{p}_{1,2}$. Combining with suitable reciprocal lattice vectors, the total momentum of two spinons with momentum $\mathbf{p}_i$ and $\mathbf{p}_j$ correspond to momentum of $\Gamma$ and $M$ points. % The prediction can be compared with numerical calculations as well as experimental measurements.
It should be borne in mind that the scattering intensity
%at the $\Gamma$ point
will be reduced by the antiferromagnetic structure factor (recall that our spinons are coarse-grained degrees of freedom).

%To facilitate such comparisons, we stress two crucial points. First, our calculation for the edge of two-spinon continuum is only valid close to $M$ and $\Gamma$ points. For higher energy transfers, other singlets that are incorporated as constraints on spinon motion can be broken and the variational approach we take stops being valid. Second, the shortest singlet two spinons can form on kagome is a nearest-neighbor singlet. Since we construct our model on the honeycomb lattice, the spinons can form onsite singlets which has zero size. To remedy this, the structure factor obtained in numerics or experiments should be divided by the form factor of a nearest-neighbor singlet with random directions on kagome before comparing with calculations from our model.

To understand the nature of the ``superconducting'' phase, we integrate out the fermions. If we neglect the amplitude fluctuations of $\Delta$, the low energy theory can be described as compact $U(1)$ gauge field interacting with a charge $2$ Higgs field. The Euclidean action for the theory in $d = 2+1$ is \cite{PhysRevD.19.3682,Bhanot1981357}
\begin{equation}\label{eq:lowenergyaction}
S=\beta\sum_{\mathbf{r}^\prime}\cos \theta(\mathbf{r}^\prime)+h\sum_{\mathbf{r},\mu}\cos(\delta_{\mu}\phi(\mathbf{r})-Q\theta_{\mu}(\mathbf{r})).
\end{equation}
Here the summation of $\mathbf{r}^\prime$ goes over all plaquettes. $\theta(\mathbf{r}^\prime)$ is the flux through plaquette $\mathbf{r}^\prime$. $\phi(\mathbf{r})$ is the phase of the Higgs field at site $\mathbf{r}$. $Q=2$ is the charge of Higgs field. $\theta_{\mu}(\mathbf{r})$ is $\phi_{\mathbf{r},\mathbf{r}+\hat{\mu}}$ in our language. $\delta_{\mu}$ is the difference operator defined as:
\begin{equation}\label{eq:difference}
\delta_{\mu}f(\mathbf{r})=(-1)^{\mathbf{r}}(f(\mathbf{r}+\tau_{\mathbf{r}}\hat{\mu})-f(\mathbf{r})).
\end{equation}
A Monte-Carlo study \cite{Bhanot1981357} of model \eqref{eq:lowenergyaction} determined its phase diagram. At low temperature ($\beta,h\to \infty$ with $\beta/h$ fixed) the theory enters a Higgs phase which preserves local $Z_2$ gauge symmetry. In other words, the ground state of the theory is a $Z_2$ liquid.

We estimate $\beta/h\sim t^2/\Delta^2\gg 1$. Based on the phase diagram (Figure $5$ of Bhanot \emph{et al.} \cite{Bhanot1981357}), we expect the finite temperature phase transition into the $Z_2$ liquid phase belongs to the three-dimensional XY universality class. The transition temperature is determined by the spin gap. The conclusion does \emph{not} depend on our saddle-point choice of $\pi$-flux phase. It is only based on that $t\sim J$ is much larger than the spin gap.

\section{$U\gg t$ limit: limitations of the model}
In the $U\to \infty$ limit, all dimer covering states have the same energy. Using these states as our basis, we employ the degenerate perturbation theory to calculate the energy contribution of the hopping term to the lowest order in $t$, $t^6/U^5$ in this case.

Without explicitly calculating these terms, their general features already reveal the limitation of our model. These operators shift dimers around a star collectively. The amplitudes of the shifts is:
\begin{equation}\label{eq:amplitudes}
-\frac{t^6}{U^5}c(h)
\end{equation}
where $c(h)>0$ is the symmetry factor depending on the number of dimers and their configurations.

These amplitudes have been derived from $S=1/2$ Heisenberg antiferromagnetic model using the overlap expansion. Our results do \emph{not} match such calculations. In other words, our model is not expected to give a good descriptions of singlet excitations.

\section{Discussion}
We first comment on connections between our model and the slave-fermion theories, \cite{PhysRevB.63.014413,PhysRevLett.98.117205,PhysRevB.77.224413,PhysRevB.83.224413} particularly the algebraic spin liquid.\cite{PhysRevLett.98.117205,PhysRevB.77.224413} Like the previous authors, we focus on a phase with a U(1) flux $\pi$ on hexagonal plaquettes and find, for a similar gauge choice, low-energy fermion excitations near momenta $\mathbf{p}_{1,2} = \pm (\pi/2, \pi/2\sqrt{3})$. Furthermore, the four bands in our model have the same dispersions (aside from an overall scale factor) as the four upper bands \cite{PhysRevB.63.014413} of the algebraic spin liquid (Appendix \ref{slave}). In this sense, it seems that our theory is an economical low-energy description of the slave fermion theory. The additional two bands in the slave-fermion models are dispersionless and could be related to spinons of another flavor,\cite{PhysRevLett.103.187203} which also happen to be dispersionless in the pure Heisenberg model.

In our approach, the compact $U(1)$ gauge theory emerges as a natural way to incorporate the constraint of dimers on spinon motion. In contrast, the gauge field is used to enforce the constraint on spin length, which translates into a requirement of one spinon per site. All states in our variational basis satisfy this constraint.

%However, such an assertion does \emph{not} have any concrete mathematical basis. In fact, there are many differences between the two theories. The fermionic statistics for our spinons is a consequence of phase ambiguity of the dimer wave function while it was built in for the slave fermion formalism. In our model, the compact $U(1)$ gauge theory arises as the natural way to incorporate the constraint of dimers on spinon motion while it is a tool to enforce the one particle per site constraint for slave fermions. Last but not least, our spinons experience strong pair-wise attraction since they can lower their energies by forming nearest neighbor singlets. The attraction is crucial for the model to host a $Z_2$ liquid ground state.  Such an interaction is not present, as far as we know, in the slave fermion formulation.

Our model shows that low-energy triplet excitations exist at both $\Gamma$ and $M$ points. It also predicts the universality class of the quantum phase transition into the $Z_2$ liquid phase. In real compounds, the presence of the DM interaction and of magnetic impurities will affect the two-spinon edge. While comparing with future measurements on more ``ideal'' materials is certainly desired, it is crucial to carry out calculations taking into account such perturbations, which will be the focus of the future work.

\textit{Acknowledgement}: We acknowledge helpful discussions with A. A. Burkov, M. J. P. Gingras, S. S. Lee, and Y. Wan. ZH was supported by NSERC of Canada. OT was supported by the U.S. Department of Energy, Office of Basic Energy Sciences, Division of Materials Sciences and Engineering under Award DEFG02-08ER46544.

\appendix
\section{The motion of a spinon on the sawtooth chain}\label{saw}
We consider a spinon on an one-dimensional chain of corner sharing triangles, the saw tooth chain \cite{PhysRevB.53.6401,PhysRevB.53.6393}. Its equation of motion is derived using both bosonic and fermionic conventions of dimer wave functions.
\subsection{The bosonic convention}
We define the state with a spinon on site $2n$ to be:
\begin{equation}\label{eq:nstate}
|n,\sigma\rangle_b=b_{2n,\sigma}^\dagger|0\rangle\prod_{j<n}|2j,2j+1\rangle\prod_{j>n}|2j,2j-1\rangle
\end{equation}
where $b_{2n,\sigma}^\dagger$ creates a bosonic spinon of spin $\sigma$ at site $2n$.

The exchange interaction for bond $\langle ij\rangle$ can be written as:
\begin{equation}\label{eq:exchangeij}
H_{ij}=\frac{J}{2}P_{ij}-\frac{J}{4}
\end{equation}
where $P_{ij}$ permutes the spin state of site $i$ and $j$. Applying $H_{2n,2n+1}$ to $|n,\sigma\rangle_b$, we get:
\begin{equation}\label{eq:move}
H_{2n,2n+1}|n\rangle_b=-\frac{J}{2}|n+1\rangle_b+\frac{J}{4}|n\rangle_b.
\end{equation}
 The full equation of motion reads \cite{PhysRevLett.103.187203,PhysRevB.81.214445}:
 \begin{equation}\label{eq:eofmboson}
 H|n\rangle_b=-\frac{J}{2}(|n+1\rangle_b+|n-1\rangle_b)+\frac{5J}{4}|n\rangle_b.
 \end{equation}

 \subsection{The fermonic convention}
 We again define:
 \begin{equation}\label{eq:f2n}
 |n,\sigma\rangle_f=a_{2n,\sigma}^\dagger |0\rangle\prod_{j<n}|2j,2j+1\rangle_f\prod_{j>n}|2j,2j-1\rangle_f.
 \end{equation}

The full equation of motion can be derived similarly:
 \begin{equation}\label{eq:eofmfermion}
 H|n\rangle_f=\frac{J}{2}(|n+1\rangle_f+|n-1\rangle_f)+\frac{5J}{4}|n\rangle_f.
 \end{equation}

The spinon hopping amplitude changes sign comparing to equation \ref{eq:eofmboson}. Similar sign-change persists on both the Husimi Cactus and the kagome lattice. In our model, spinons live on the honeycomb lattice with plaquettes of even length. The overall sign of spinon hopping amplitude is thus irrelevant since it can be reversed by a simple gauge transformation.

\section{Slave-fermion mean-field theory on kagome}\label{slave}
In this section, we solve for the mean-field band structure  of the algebraic spin liquid. In the slave fermion theory, a spin operator $S_{i}^{(a)}$ is written in terms of fermions:
\begin{equation}\label{eq:sslave}
S_{i}^{(a)}=\frac{1}{2}a_{i\sigma}^\dagger \sigma_{\sigma\sigma^\prime}^{(a)}a_{i\sigma^\prime}
\end{equation}
where $\sigma^{(a)}$ with $a=x,y,z$ are the Pauli matrices.

The exchange Hamiltonian can be written as:
\begin{equation}\label{eq:eslave}
H=\frac{J}{2}\sum_{\langle ij\rangle} a_{i\sigma}^\dagger a_{i\sigma^\prime} a_{j\sigma^\prime}^\dagger a_{j\sigma}.
\end{equation}
We decompose the four-fermion terms using mean-field approximation \cite{PhysRevB.63.014413}:
\begin{equation}\label{eq:mfslave}
H_\mathrm{MF}\approx
	\frac{J}{2}\sum_{\langle ij\rangle}(\chi_{ij}a_{j\sigma}^\dagger a_{i\sigma} + \mathrm{H.c.})
	+\frac{J}{2}\sum_{\langle ij\rangle}|\chi_{ij}|^2,
\end{equation}
with the self-consistency condition $\chi_{ij} = -\langle a_{i\sigma}^\dagger a_{j\sigma}\rangle$. By symmetry, $|\chi_{ij}| = \chi$ for every nearest-neighbor bond $\langle ij\rangle$. Assuming that fluxes through hexagons and triangles are $\pi$ and 0, respectively, we obtain six bands, in the order of increasing energy:
\begin{subequations}\label{eq:6bands}
\begin{eqnarray}
\xi_{1,2}(\mathbf{k})&=&-J\chi, \\
\xi_{3,4}(\mathbf{k})&=&\frac{J\chi}{2}(1-\sqrt{3\pm2\rho_{\mathbf{k}}}),\\
\xi_{5,6}(\mathbf{k})&=&\frac{J\chi}{2}(1+\sqrt{3\mp2\rho_{\mathbf{k}}}).
\end{eqnarray}
\end{subequations}
These are shown in the right panel of Fig.~\ref{fig:bandpi}.
\bibliography{z2liquid,thesis}
\end{document}